\documentclass[aps,prl,twocolumn,groupedaddress,showpacs,showkeys]{revtex4}
\usepackage{graphicx}
\begin{document}
\title{On ultrafast magnetic flux dendrite propagation into thin
superconducting films}
\author{B. Biehler}
\affiliation{Physics Department, University of Konstanz, D-78457
Konstanz, Germany}
\author{B.-U. Runge}
\affiliation{Physics Department, University of Konstanz, D-78457
Konstanz, Germany}
\author{P.~Leiderer}
\affiliation{Physics Department, University of Konstanz, D-78457
Konstanz, Germany}
\author{R. G. Mints}
\email[]{mints@post.tau.ac.il}
\affiliation{School of Physics and Astronomy, Raymond and Beverly
Sackler Faculty of Exact Sciences, Tel Aviv University, Tel Aviv 69978,
Israel}
\date{\today}
\begin{abstract}
We suggest a new theoretical approach describing the velocity of
magnetic flux dendrite penetration into thin superconducting films. The
key assumptions for this approach are based upon experimental
observations. We treat a dendrite tip motion as a propagating flux jump
instability. Two different regimes of dendrite propagation are found: A
fast initial stage is followed by a slow stage, which sets in as soon
as a dendrite enters into the vortex-free region. We find that the
dendrite velocity is inversely proportional to the sample thickness.
The theoretical results and experimental data obtained by a
magneto-optic pump-probe technique are compared and excellent agreement
between the calculations and measurements is found.
\end{abstract}
\maketitle
Magnetic flux penetration in type-II superconductors is successfully
described by the Bean critical state model \cite{Bea62}. This model
assumes that the slope of the flux ``hills'' is given by
$\mu_0j_c(T,B)$, where the critical current density $j_c(T,B)$ is a
decreasing function of the temperature $T$ and field $B$. Bean's
critical state with its spatially nonuniform flux distribution is not
at equilibrium and under certain conditions the smooth flux penetration
process becomes unstable (see review \cite{Min81} and references
therein). The spatial and temporal development of this instability
depends on the sample geometry, temperature, external magnetic field,
its rate of change and orientation, initial and boundary conditions,
etc.
\par
Instabilities in the critical state result in flux redistribution
towards the equilibrium state (spatially homogeneous flux throughout
the sample) and are accompanied by a significant heat release, which
often leads to the superconductor-to-normal-transition. The basic
instability observed in Bean's critical state is the flux jump
instability, which was discovered already in the early experiments on
superconductors with strong pinning \cite{Min81}.
\par
The basic physics of flux jumping can be easily illustrated. Assume a
perturbation of temperature or flux occurring in Bean's critical state.
This perturbation can be caused by an external reason or a spontaneous
fluctuation arising in the system itself. The initial perturbation
redistributes the magnetic flux inside the superconductor. This flux
motion by itself induces an electric field which leads to dissipation,
since the electric field does not only act on the Cooper pairs but also
on the unpaired electrons. This additional dissipation results in an
extra heating which in turn leads to an additional flux motion. This
``loop'' establishes a positive feedback driving the system towards the
equilibrium state. The flux jumping instability exhibits itself as
suddenly appearing flux avalanche (flux jump) and heat release
\cite{Min81,Alt04}.
\par
Spatially resolved flux front patterns of Bean's critical state
instability were first observed in Nb discs with thicknesses in the
range of $d\approx 10^{-5}\,$m to $10^{-3}\,$m by means of
magneto-optic imaging \cite{Wer67}. Wertheimer and Gilchrist discovered
a well defined pattern of flux dendrites with a width $w\sim
10^{-3}\,$m and propagation velocity $v$ in the interval between
$5\,$m/s and $100\,$m/s \cite{Wer67}. The dendrites velocity depended
on the disks thickness, for smaller $d$ a higher $v$ was found.
\par
The modern magneto-optic technique allowing to investigate flux
patterns with time resolution on the order of $\approx 100\,$ps
\cite{Fre92,Fre91} stimulated quite a few experimental and theoretical
studies of flux front patterns arising in a process of smooth flux
penetration \cite{Bra95} as well as in a process of critical state
instability development in superconducting films in a transversal
magnetic field. Different scenarios are considered resulting in a
variety of flux patterns, e.g., magnetic turbulence \cite{Vla94,Kob98},
kinetic flux front roughening \cite{Sur99}, magnetic micro avalanches
\cite{Fie95,Now97}, flux dendrites
\cite{Lei93,Bol03a,Dur95,Ara01,Ara04}, thermomagnetic fingering
\cite{Rak04}, bending of flux-antiflux interface \cite{Bas98a,Fis04},
and flux front corrugation \cite{Bas98b}.
\par
A wealth of recent experiments convincingly demonstrate that a
propagating dendritic flux pattern driven by the flux jumping
instability is a general phenomenon typical for Bean's-type critical
state \cite{Lei93,Dur95,Bol03a,Buj93,Bol03b,Rud03,Joh02}. Indeed, the
flux dendrites were observed under a wide variety of conditions in
superconducting films of Nb \cite{Lei93,Dur95,Buj93},
YBa$_2$Cu$_3$O$_{7-\delta}$ \cite{Lei93,Bol03a,Bol03b}, Nb$_3$Sn
\cite{Rud03}, and MgB$_2$ \cite{Joh02}.
\par
It is known, that dendrite propagation in thin films shows velocities
up to 160$\,$km/s \cite{Bol03a}, {\em i.e.}, these velocities are much
higher than the speed of sound. This ultrafast motion of flux dendrites
in thin superconducting films is a long standing and challenging
problem.
\par
In this letter we derive a novel equation for a dendrite tip velocity
and demonstrate an excellent agreement between the theoretical results
and experimental data for the propagation velocity of a single flux
dendrite branch.
\par
%%%%%%%%%%%%%%%%%%%%%%%%   FIGURE 1  %%%%%%%%%%%%%%%%%%%%%%%%%
\begin{figure}[t]
\includegraphics[width=0.95\hsize]{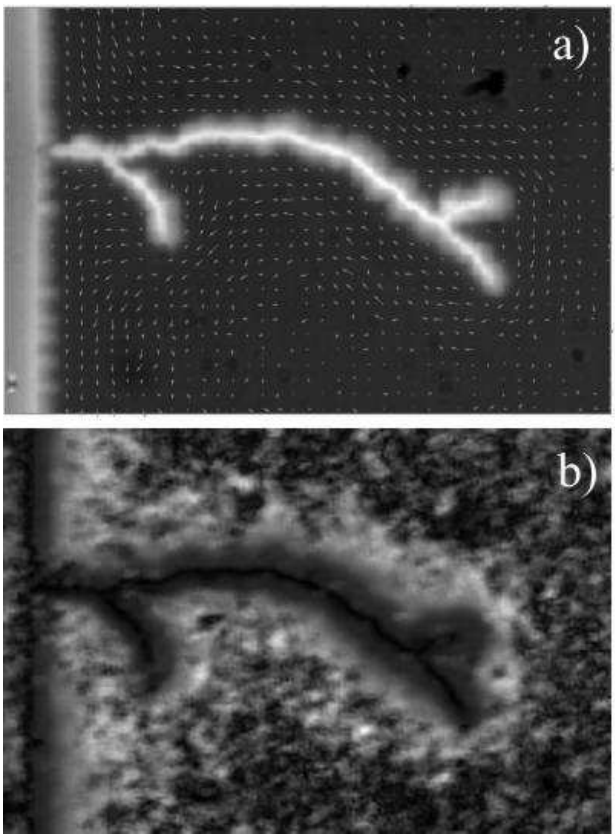}
\caption{Magneto-optic images of a dendritic flux pattern in a YBCO
film with the thickness $d=330\,$nm subjected to a field of
$B_a=15\,$mT. (a) Final state (after $\approx 10\,$s) of a dendritic
flux pattern with superimposed current distribution shown by the
arrows. The length of the arrows is proportional to the local current
density. (b) The absolute value of the current density is shown. The
bright areas indicate high current densities.
\label{fig_1}}
\end{figure}
%%%%%%%%%%%%%%%%%%%%%%%%%%%%%%%%%%%%%%%%%%%%%%%%%%%%%%%%%%%%%%%
\par
Dendritic flux structures which can be considered as a set of single
flux branches originating from a certain area were observed in numerous
experiments \cite{Bol03a}. In the case of a dendritic structure with
few branches the single branches do not affect each other and the
propagating substructures can be treated as a moving flux jump
instability localized at the tip of the dendrite branches. A typical
magneto-optic image of a ``dilute'' dendrite in its final state is
shown in Fig.~\ref{fig_1}(a). Superimposed are the current streamlines
as determined by an inversion scheme \cite{Wij98}. In
Fig.~\ref{fig_1}(b) the absolute value of the current density is shown.
It is worth mentioning that the center of the dendrite is current free
and that the current follows the dendrite branches. The current density
decreases rapidly with distance from the dendritic structure.
\par
These experimental observations allow for a straight-line flux dendrite
model, which we use in our calculations. This model assumes the
following:
\par
\begin{figure}[b]
\includegraphics[width=0.94\hsize]{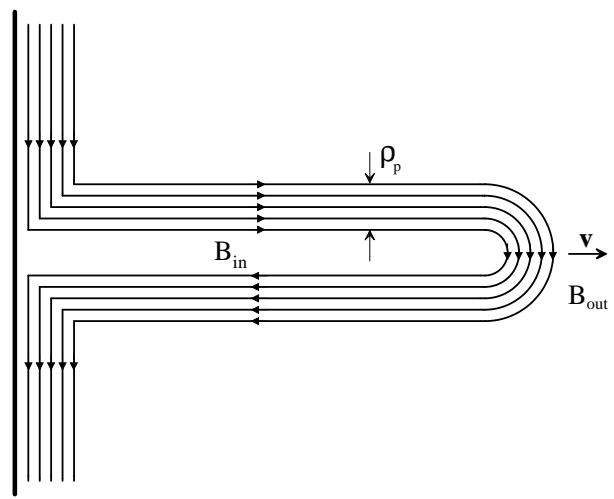}
\caption{Current lines for the straight-line magnetic flux dendrite.
The full line to the left marks the strip edge.
\label{fig_2}}
\end{figure}

(a) The current of a straight-line dendrite first flows parallel to the
sample edge, then closely follows the contour of the dendrite branch
until flowing parallel to the sample edge again as shown in
Fig.~\ref{fig_2}.
\par
(b) In the current carrying areas, the superconductor is in the flux
creep regime and thus the current density $j$ depends on the electric
field $E$ as
% 1
\begin{equation}\label{eq1}
j=j_c\left({E/E_0}\right)^{1/n},
\end{equation}
where $j_c$ is the critical current density, $n$ and $E_0$ are the
parameters characterizing the current density-electric field curve (at
$E=E_0$ we have $j=j_c$). It is common to define $j_c$ as the current
density at $E_0=10^{-4}\,$V/m, for high-$T_c$ superconductors $n\sim
10$ but decreases with the applied magnetic field \cite{Eki89}.
Eq.~(\ref{eq1}) yields the electric field dependent conductivity
% 2
\begin{equation}\label{eq2}
\sigma (E)={dj\over dE}\approx {j_c\over nE}.
\end{equation}
\par
%%%%%%%%%%%%%%%%%%%%%%%%   FIGURE 2   %%%%%%%%%%%%%%%%%%%%%%%%%
%%%%%%%%%%%%%%%%%%%%%%%%%%%%%%%%%%%%%%%%%%%%%%%%%%%%%%%%%%%%%%%
\par
Next, we denote the radius of the dendrite tip as $\rho_0$ and the
width of the current carrying area as
\par
% 3
\begin{equation}\label{eq3}
\rho_p={B_{\rm eff}/\mu_0 j_c},
\end{equation}
where $B_{\rm eff}=B_{\rm in} - B_{\rm out}$, $B_{\rm in}$ is the
field inside the dendrite, and $B_{\rm out}$ is the field outside
the tip of the dendrite.
\par
Consider now the flux front stability at a tip of a moving flux
dendrite in the framework of the model developed to treat the flux jump
instability near a semicircle indentation at the sample edge
\cite{Min82,Min96a,Min96b}. This approach is based on the assumption
that the flux jumping instability develops much faster than the
magnetic flux diffusion. In the flux creep regime of low-$T_c$ and
high-$T_c$ superconductors this assumption holds with a high accuracy
\cite{Min82,Min96a}.
\par
It follows from the general approach that the stability margin of a
flux jumping instability is determined by the existence of a nontrivial
solution of the equation \cite{Min96a,Min96b}
% 4
\begin{equation}\label{eq4}
\Delta\theta-q^2\theta+ {nE\over\lambda}\, \Big|{\partial
j_c\over\partial T}\Big|\,\theta=0,
\end{equation}
where $\theta$ is the temperature perturbation, $\lambda$ is the heat
conductivity, $E$ is the electric field generated by a time dependent
magnetic field, the parameter $q$ is given by
% 5
\begin{equation}\label{eq5}
\tan qd={h/\lambda q},
\end{equation}
and $h$ is the heat transfer coefficient to the coolant. The boundary
condition to Eq.~(\ref{eq4}) is ${\bf n}\nabla\theta=0$ at the edge of
the film and $\bf n$ is the unit vector perpendicular to the edge of
the film. It is clear from Eq.~(\ref{eq4}) that the flux front
stability is highly sensitive to the electric field $E$ generated by
the varying magnetic field \cite{Min82,Min96a,Min96b}.
\par
The dendrite tip motion results in an electric field ${\bf E}$, which
is parallel to the current density ${\bf j}$. We consider this field
similar to the consideration of the electric field generated by a
varying magnetic field at a semicircular indentation with a radius
$\rho_0$ in a superconducting film with a straight edge \cite{Min96b}.
This approach results in
% 6
\begin{equation}\label{eq6}
E\approx \dot{B}_{\rm in}\,\rho_p^2/\rho_0 \, .
\end{equation}
Assuming that $\rho_0\le\rho_p$ we estimate the magnetic field rate in
the vicinity of a dendrite tip as
% 7
\begin{equation}\label{eq7}
\dot{B}_{\rm in}\approx vB_{\rm eff}/\rho_p\,.
\end{equation}
Combining the Eq.~\ref{eq6} and~\ref{eq7} we find that the electric
field generated at the inner edge of a moving flux dendrite tip can be
estimated as $E\approx vB_{\rm eff}\rho_p/\rho_0$.
\par
Next, we assume a high thermal boundary resistance in the films, which
means that $hd\ll\lambda$. This assumption can be justified using
typical values $\lambda =0.1\,$W$\,$K$^{-1}\,$m$^{-1}$, $d=330\,$nm,
and $h=(700 \textrm{\ to}\ 1.3\times 10^3)\,$W$\,$K$^{-1}\,$m$^{-2}$
\cite{Coh92, Nah91}. In this case of $hd\ll\lambda$ we find from Eq.
(\ref{eq5}) that the value of $q^2\approx h/\lambda d$ and the
stability criterion for flux jumping in thin film takes the
form\cite{Min96b}
% 8
\begin{equation}\label{eq8}
{B_{\rm eff}^2\dot{B}_{\rm in}nd\over 2\mu_0^2\rho_0 hj_c^2}\,
\left|{dj_c\over dT}\right|=1\,.
\end{equation}
\par
Consider now the critical current density to be linear in temperature,
{\it i.e.}, $j_c=j_0(1-T/T_c)$. In this case we use  Eq. (\ref{eq7}) to
rewrite Eq. (\ref{eq8}) as follows
% 9
\begin{equation}\label{eq9}
v=\gamma\,{2\mu_0^2j_c^2hT_c\rho_0\rho_p \over ndj_0B_{\rm eff}^3}\,,
\end{equation}
where $\gamma\sim 1$ is a numerical factor and $j_0$ is the critical
current density at $T=0$. The criterion given by Eq.~(\ref{eq9}) gives
the lower limit of a dendrite tip speed. If we assume that
$\rho_0\approx\rho_p\,$ and use Eq.~(\ref{eq3}) then
% 10
\begin{equation}\label{eq10}
v=2\gamma\,{hT_c\over ndB_{\rm eff}j_0}\,= 2\gamma\,{hT_c\over
nd(B_{\rm in}-B_{\rm out})j_0}\,.
\end{equation}
\par
We compare now the results obtained by Eq. (\ref{eq10}) and our
experimental data. To measure the time dependent dendrite length
$s=s(t)$ we used a magneto-optic pump-probe setup \cite{Bol03a}. The
dendrites where nucleated at the edge of a square YBCO thin film sample
by focusing a laser beam onto the film surface. The magnetic field was
applied prior to the laser pulse. We observed two qualitatively
different stages of dendrite propagation. In the first few nanoseconds
we observed an extremely high velocity on the order of 160$\,$km/s,
later on this velocity decreased to a value of 18$\,$km/s. For
experimental details see Ref.~\cite{Bol03a}. The existence of these two
distinct regions of dendrites propagation can be easily understood
using Eq.~(\ref{eq10}). Indeed, as long as a dendrite crosses the
critical state area the field $B_{\rm out}$ is decreasing, therefore
the value of $B_{\rm eff}=B_{\rm in}-B_{\rm out}$ is increasing and
consequently the velocity of the dendrite is decreasing. After the
dendrite tip crosses the critical state area its velocity stays
constant as the dendrite runs in a vortex free area where $B_{\rm eff}$
is a constant.
\par
%%%%%%%%%%%%%%%%%%%%%%%%   FIGURE 3   %%%%%%%%%%%%%%%%%%%%%%%%%
\begin{figure}[t]
\includegraphics[width=.94\hsize]{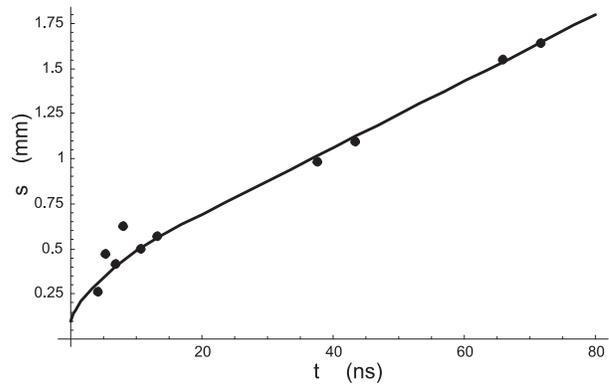}
\caption{The time dependence of the length of a magnetic flux dendrite
$s(t)$. The solid line is the solution of Eq.~(\ref{eq12}) and the
solid dots are the experimental data \cite{Bol03a}.}
\label{fig_3}
\end{figure}
%%%%%%%%%%%%%%%%%%%%%%%%%%%%%%%%%%%%%%%%%%%%%%%%%%%%%%%%%%%%%%%
\par
The time dependence of the dendrite length $s=s(t)$ can be calculated
using Eq.~(\ref{eq10}). The effective field $B_{\rm eff}$ is the
crucial parameter for this calculation. To find $B_{\rm eff}$ we assume
that the magnetic field inside a long superconducting strip is a good
approximation for the distribution at the center of a square
superconducting thin film \cite{Bra93}
% 11
\begin{equation}
\label{eq11}
B_{\rm out}(s)={\mu_0 j_0\over\pi}
\left\{
\begin{array}{l@{\quad}l@{\ \ }r}
0,
&|s|<b\\
\ \\
\textrm{artanh}\Big(\displaystyle{\sqrt{s^2-b^2}\over c|s|}\Big),
&b<|s|<a\\
\ \\
\textrm{artanh}\Big(\displaystyle{c|s|\over \sqrt{s^2-b^2}}\Big),
&|s|>a
\end{array}
\right.
\end{equation}
where $a$ and $b$ are the half widths of the superconductor and of the
vortex free area, and $c=\sqrt{1-b^2/a^2}$. Substituting
Eq.~(\ref{eq11}) into Eq.~(\ref{eq10}) we obtain a differential equation
for the dendrite propagation
% 12
\begin{equation}
\label{eq12}
\frac{ds}{dt} = \frac{\alpha}{B_{\rm in}-B_{\rm out}(s)},
\end{equation}
with $\alpha = 2 \gamma h T_c/ndj_0$. Based on our experimental data we
assume that the field $B_{\rm in}$ is constant and by a factor of 1.9
larger than the applied magnetic field.
\par
A numerical solution of Eq.~(\ref{eq12}) yields the solid line in
Fig.~\ref{fig_3}. We used for this plot the values $d=330\,$nm,
$a=5\,$mm, $b=4.4\,$mm, $T_c=90\,$K, $j_0=1.5\times 10^9\,$A/m$^2$,
$h=10^4\,$W$\,$K$^{-1}$m$^{-2}$, $T_c=90\,$K, $\gamma =1$, and $n=6$.
We take $s(0) = a-\ell$ with $\ell =0.1\,$mm as an initial condition to
avoid the singularity of $B_{\rm out}(s)$ at $t=0$.
\par
%%%%%%%%%%%%%%%%%%%%%%%%   FIGURE 4   %%%%%%%%%%%%%%%%%%%%%%%%%
\begin{figure}[t]
\includegraphics[width=.94\hsize]{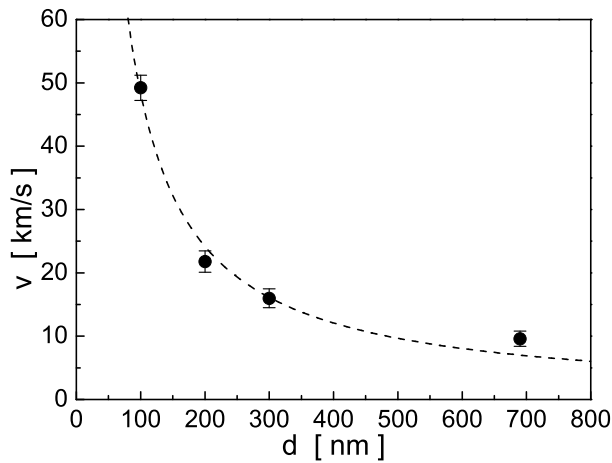}
\caption{The dependence of the flux dendrite velocity on the sample
thickness $v(d)$. The dashed line is a fit of $v \propto 1/d$ revealing
the dependence given by Eq. (\ref{eq7}) and the solid dots are the
experimental data \cite{Bol02}.}
\label{fig_4}
\end{figure}
%%%%%%%%%%%%%%%%%%%%%%%%%%%%%%%%%%%%%%%%%%%%%%%%%%%%%%%%%%%%%%%
\par
To check Eq.~(\ref{eq10}) further we compare the calculated velocities
with the velocities obtained from line-focus measurements
\cite{Bol03a}. If a line focus is applied, the dendrites never run in
the critical state, but penetrate into flux free area. In this case, as
expected from Eq.~(\ref{eq10}), we don't find a regime with increased
velocities, however we find a thickness dependence. In Fig.~\ref{fig_4}
one can see the experimentally obtained thickness dependent velocity
and a fit $v \propto 1/d$. One reason for the slight deviations between
the theory and experiment may be that we had to use different YBCO
films to obtain the data, {\it i.e.}, the values for parameters like
$T_c$ or $B_{\rm eff}$ may vary.
\par
The main result of this letter is Eq.~(\ref{eq10}). It describes the
dynamics of a single flux dendrite and it was shown, that an excellent
agreement with the experimental data has been achieved. To the authors'
knowledge this is the first time that a theory explains the
observation of the fast and slow penetration velocities.
\par
\begin{acknowledgments}
We want to thank the Kurt Lion Foundation, the Israeli Science
Foundation (grant No 283/00-11.7), and the German Israeli Foundation
(grant No G-721-150.14/10) for their support. Further, we want to thank
H.~Kinder for providing the YBCO samples.
\end{acknowledgments}
%
%**********************bibliography**************************

%
\end{document}